\documentclass[conference]{IEEEtran}

\usepackage[hyphens]{url}
\usepackage{hyperref}
\usepackage{inputenc}
\usepackage{multirow}
\usepackage{caption}
\usepackage{siunitx} 
\usepackage{float}
\begin{document}

\title{Shaping a Quantum-Resistant Future: \\Strategies for Post-Quantum PKI}

\author{\IEEEauthorblockN{Grazia D'Onghia}
\IEEEauthorblockA{Politecnico di Torino\\
Dip. di Automatica e Informatica\\
Torino, Italy\\
grazia.donghia@polito.it}
\and
\IEEEauthorblockN{Diana Gratiela Berbecaru}
\IEEEauthorblockA{Politecnico di Torino\\
Dip. di Automatica e Informatica\\
Torino, Italy\\
diana.berbecaru@polito.it}
\and
\IEEEauthorblockN{Antonio Lioy}
\IEEEauthorblockA{Politecnico di Torino\\
Dip. di Automatica e Informatica\\
Torino, Italy\\
antonio.lioy@polito.it}
}
\maketitle
\thispagestyle{plain}
\pagestyle{plain}
\begin{abstract}
As the quantum computing era approaches, securing classical cryptographic protocols becomes imperative. Public key cryptography is widely used for signature and key exchange, but it's the type of cryptography more threatened by quantum computing. Its application typically requires support via a public-key certificate, which is a signed data structure and must therefore face twice the quantum challenge: for the certified keys and for the signature itself. We present the latest developments in selecting robust Post-Quantum algorithms and investigate their applicability in the Public Key Infrastructure context. Our contribution entails defining requirements for a secure transition to a quantum-resistant Public Key Infrastructure, with a focus on adaptations for the X.509 certificate format. Additionally, we explore transitioning Certificate Revocation List and Online Certificate Status Protocol to support quantum-resistant algorithms. Through comparative analysis, we elucidate the complex transition to a quantum-resistant PKI.
\end{abstract}

\IEEEpeerreviewmaketitle

\section{Introduction} \label{introduction}
Quantum computing is expected to revolutionize computational speed and performance but threatens current cryptosystems. Shor's algorithm allows quantum computers to factor large integers and compute discrete logarithms in polynomial time, undermining asymmetric cryptography like RSA and ECDSA algorithm in use nowadays.
Grover's algorithm accelerates brute-force search, which could compromise symmetric encryption such as AES. However, symmetric encryption can be made Quantum-Safe (QS) by increasing key lengths: for example, AES-256 is QS \cite{qs_white_paper}.
Despite these risks, many organizations do not perceive an urgency to transition to QS systems \cite{quantum_safe_information_sharing}.
The delay in adopting quantum-resistant solutions could be problematic, especially for data that must remain secure for years, as quantum attackers might follow a ``store-now-decrypt-later'' strategy \cite{towards_security_recommendations_pq_pki}, collecting encrypted data today to decrypt them once quantum technology matures.

Classical cryptosystems have been built basically in two decades revolutionizing security protocols and remote transactions. However, preparing for QS systems requires immediate action because the PQ transition and implementation into real systems will require time.
In particular, given the widespread use of digital certificates issued in the frame of a Public-Key Infrastructure (PKI), intensive research work should be dedicated to defining and standardizing PQ-enabled X.509 certificates, as well as integrating support for PQC into the main PKI protocols.
Some works and guidelines have been indicated in this sense \cite{nist_migration_pqc}, but yet nowadays it is not clear how the future PQ-enabled certificates should look like
.
As indicated in \cite{BSI}, in October 2019, ITU-T published an update of the X.509v3 standard \cite{X509}, which mentions that new signature schemes must be introduced into certificates or PKIs, but it does not indicate a fixed deadline for migration.  The ITU-T concludes: ``\textit{it is unlikely that it is possible to change cryptographic algorithms \textit{simultaneously} for all
entities within a PKI}''. However, we know that the process will be gradual: in the first place, certificate extensions will have to be defined to accommodate ``alternative'' public key(s). For backward compatibility reasons,
it is recommended to mark the (new) certificate extensions
as ``non-critical'', so that applications
that are not aware of the new extensions can also check
the validity of corresponding certificates.  Nevertheless, the adaptations in the certificate structure are only intended as a ``temporary'' solution until the
migration process to (pure) QS signature schemes
is completed. In this regard, the ITU writes: ``\textit{After the
migration period, it is expected that new public-key certificates be issued without these extensions and with the new
set of cryptographic algorithms and the digital signature in
the base part of the public-key certificate.}'' 

Our study aims to fill the gap between expectations and concrete definitions of PQ-based certificates by providing a roadmap for the transition to PQ PKI. We focus on general and more specific requirements for adapting the certificate format, and then we overview some existing implementations and previously reported results. We also discuss the migration for the (currently standardized) mechanisms, namely the Certificate Revocation Lists (CRLs) and Online Certificate Status Protocol (OCSP) \cite{RFC6960} although various applications currently skip revocation verification, or adopt customized methods so perhaps in the future alternative solutions would occur. 

The paper is structured as follows: 
Section~\ref{PQCalgs} 
briefly presents the PQC algorithms and the current status of PQC standardization. The underlying cryptographic mechanisms and their quantum resistance is out of the scope of this paper, but it is essential for understanding the full implications of the proposed changes to PKI systems. 
Section~\ref{currentPKI} reviews the core PKI components, in particular, the current X.509v3 certificate and CRL formats, and OCSP.
Section \ref{pq_transition_pki} gives general requirements for transitioning to PQ PKI and discusses the challenges and design considerations for PQ-based X.509 certificates, CRLs, and OCSP. 
Finally, Section \ref{conclusions} presents conclusions and identifies potential 
future works.
\section{PQ Cryptography} \label{PQCalgs}

PQC algorithms fall into several distinct categories related to hard mathematical problems, whose detailed analysis is beyond the scope of this paper. These categories are \cite{navigating_transition}:

1. \textit{Lattice-based cryptography}: Algorithms like NTRU and ring-LWE (Learning With Errors) offer security but require longer keys than RSA.

2. \textit{Code-based cryptography}: McEliece and Courtois-Finiasz-Sendrier (CFS) require large key sizes.

3. \textit{Hash-based cryptography}: Merkle and XMSS are robust but produce large signatures.

4. \textit{Multivariate cryptography}: Rainbow algorithm offers small signatures but less efficiency.

5. \textit{Isogeny-based cryptography}: SIKE provides smaller key sizes for Diffie-Hellman-like key exchange.

Given the variety of PQ algorithms, each with unique strengths and weaknesses, it's crucial to understand the technical requirements for selecting the most appropriate ones in a QS cryptosystem. The choice of algorithm(s) depends on factors like key size, efficiency/speed, security/robustness, and intended use cases. 
Thus, we present briefly the NIST algorithms, delineating their main characteristics.

\begin{table*} 
    \centering
    \caption{NIST PQ algorithms: key sizes and security levels.}
    \begin{tabular}{|c|c|c|c|c|}
         \hline
\multicolumn{5}{|c|}{\textbf{PQ Algorithms}}\\
         \hline
        Name & Public Key size &Private Key size &Ciphertext/signature size & Security Level\\
         \hline
         Falcon 512&  \SI{897}{\byte}& \SI{1281}{\byte} &  \SI{666}{\byte}& 1\\
         \hline
         Falcon 1024&  \SI{1793}{\byte}&  \SI{2305}{\byte}&  \SI{1280}{\byte}&5\\
         \hline
         Dilithium 2& \SI{1312}{\byte} & \SI{2528}{\byte}  &\SI{2420}{\byte}  &2 \\
         \hline
         Dilithium 3&  \SI{1952}{\byte}& \SI{4000}{\byte} &  \SI{3293}{\byte}&3\\
         \hline
         Dilithium 5&  \SI{2592}{\byte}&  \SI{4864}{\byte}& \SI{4595}{\byte} &5\\
         \hline
         SLH-DSA-128& \SI{32}{\byte} &\SI{64}{\byte} &\SI{7856}{\byte} (small) &\\ 
         &&& \SI{17088}{\byte} (fast) &1\\
         \hline
         SLH-DSA-192&\SI{48}{\byte} & \SI{96}{\byte}& \SI{16224}{\byte} (small) &\\
         &&& \SI{35664}{\byte} (fast)&3\\
         \hline
         SLH-DSA-256& \SI{64}{\byte}& \SI{128}{\byte}& \SI{29792}{\byte} (small)&\\
         &&& \SI{49856}{\byte} (fast)&5\\
         \hline
    \end{tabular}
    
    \label{tab:my_label}
\end{table*}

\subsection{Current standardization status for PQ algorithms}
The US National Institute of Standards and Technology (NIST) has been leading the standardization process for PQ algorithms since 2016. NIST's selection criteria emphasize security levels, with a range from 1 (lowest) to 5 (highest), with symmetric encryption serving as the reference point for quantum robustness \cite{cloudflare}. Level 1 is equivalent to the security of AES-128, while level 5 is equivalent to AES-256, which is currently considered QS. Since symmetric encryption can already achieve security level 5, the same level should be achieved with PQ algorithms for digital signatures.

In 2023, NIST announced that it would standardize four algorithms, one key encapsulation mechanism (KEM), and three digital signature schemes \cite{nist_standard}:

- \textit{Kyber}: A lattice-based KEM with a good balance of security and efficiency. It provides relatively smaller key sizes, which is advantageous for various applications.

- \textit{Dilithium} and \textit{Falcon}: lattice-based digital signatures with high security and efficient verification.

-  \textit{SPHINCS+}: a hash-based signature scheme, offering robust security but larger signature sizes. SLH-DSA is a SPHINCS+ Stateless Hash-based Digital Signature algorithm, which has different schemas (SLH-DSA-128, SLH-DSA-192, and SLH-DSA-256). Hash-based signature schemes are typically considered the most robust ones, although they generate larger signatures, affecting storage and bandwidth requirements.

Table \ref{tab:my_label} summarizes the key sizes and security levels for the selected PQ algorithms. Since the PQ algorithms require large key sizes and produce bigger signatures, it is challenging to integrate them into the current PKI to maintain the current efficiency of the infrastructure. 



\section{Current PKI} \label{currentPKI}
In this section, we review the current structure of 
X.509v3 certificate and CRL, and the OCSP. 

\subsection{Current X.509v3 certificate format} \label{currentX.509}

Certificates are data structures that bind
public keys to entities (persons, end-nodes) that are issued and signed by a trusted third party, the Certification Authority (CA). 


The first element of an X.509v3 certificate is \texttt{tbsCertificate}, which is signed and contains all the necessary information to describe an entity. The cryptographic information of the certificate is embedded in the \textit{Subject Public Key Info} field, which is mandatory and contains the entity's public key.
%
The digital signature algorithm is described by the \texttt{signatureAlgorithm} element and identified by \texttt{AlgorithmIdentifier} or Object Identifier (OID).
The digital signature is then appended at the end of the certificate as a bit string of type \texttt{TBSCertificate}.
%
The (main) fields of the certificate that most probably will be affected by PQ transition are given below: 

 \begin{itemize}
     \item \textit{Version} indicates the version of the certificate, currently the version in use is v3 and is identified with an integer. 
 \item \textit{Signature} indicates the algorithm identifier (that is, the OID, plus any associated parameters) of the algorithm used to calculate the digital signature on the  certificate. 
 \item \textit{Validity} indicates the window of time or validity period that this certificate should be
 considered valid. This field is composed of two dates, the \textit{Not Valid Before} and the \textit{Not
 Valid After}. These dates/times may be represented in UTC Time or Generalized Time. There is a tendency to shorten the certificate lifetime, especially for server certificates.
 \item \textit{Subject Public Key Info} is the public key identified with a sequence containing an OID, associated with the subject. 
 \end{itemize}

Moreover, the extensions will also be affected by PQ transition, including the ones indicated below:

 \begin{itemize}
 \item \textit{Authority Key Identifier} (AKI) is the hash of the issuer's public key, calculated with  SHA-1 algorithm. It identifies the public key because the issuer may have multiple keys, then this extension allows  applications to identify the public key of the issuer, which must be used to verify the
 signature.  
     \item \textit{Subject key identifier} (SKI) is the hash value of the key that corresponds to the certificate.
 It is used by applications to compare the public key in the certificate with other public
 keys, and it is useful if the owner of the certificate has several public keys. Similar to the AKI extension, it is calculated by means of SHA1 hash function.
 \item \textit{Key Usage} indicates what the public key of the certificate can be used for, for example,
 for Digital Signatures or Non-Repudiation, among others.
 \item \textit{Extended Key Usage}:  indicates some additional usages to the key usage field, such as
 time stamping or revocation status signing.
 \end{itemize}
 Some extensions should not be affected by the PQ transition, such as \textit{certificatePolicies} and \textit{Policy Mappings}, or \textit{BasicConstraints}. 
 On the other hand, new extensions will contain alternative PQ keys, algorithms and signatures, as described in Sect. \ref{design}.
%

\subsection{CRL and OCSP}
A Certificate Revocation List (CRL), defined in \cite{RFC5280} together with the X.509 certificates is a signed data structure containing a list of revoked certificates.
A digital signature is applied by the CA (or delegated authority) on the CRL to provide integrity and authenticity.
The CRL can be downloaded (when needed) to check the revocation status of a certificate of interest. It also can be kept locally and cached to be read without having access to the Internet, which could be useful in some cases. 
The signature of CRL can be done by the same issuer CA that issued the certificate, or by a delegated entity. The current CRL version widely used is version 2 which exploits also extensions much the same as X.509 certificates. Below we provide a list of the fields affected by PQ transition:
\begin{itemize}
    
\item \textit{Version} contains the version of the encoded CRL. 
\item \textit{Signature} contains digital signature algorithm used to calculate the signature of the CRL as well as specifying the hash algorithm, for example, RSA2048 with SHA256.

\item \textit{Extensions} allows more information to be added with each revocation.
Similar to the certificate extensions, Authority Key Identifiers will be affected by PQ transition.

\end{itemize}


With the periodic CRL mechanisms, it is possible to obtain updated revocation information at
fixed points in time, typically every day or once a month. Since CRLs can be stored in the cache, it is thus possible to obtain revocation information even when the relying party is offine. However, the CRLs 
can become quite big, and consulting them can become time-consuming, impacting thus the usability of the application relying on them. Additionally, storing
them may need a lot of space which is not available for example in mobile devices. Furthermore,
the validity period and the publication of an updated CRL can take days, so the revocation information (for a checked certificate) may not be the most recent one.

The Online Certificate Status Protocol (OCSP) documented in \cite{RFC6960} allows clients to query an OCSP server about the revocation status of individual certificates. 
The main advantage is that it returns more up-to-date information than a CRL.  OCSP's main advantage is that it does not require much storage, but on the other hand, OCSP requires applications to be online.

The OCSP protocol works as follows: an OCSP client sends a request to an OCSP
server about the revocation status of one or more certificates, optionally the request can be
digitally signed. For each certificate, the request contains the serial number of the certificate along with the hash values of the issuer's DN and of the issuer's public key. This information
determines the certificate uniquely. 
The OCSP server responds to the request with a digitally signed message. If the
status of more than one certificate is requested, then the answers contain information about
the status of each requested certificate. The returned revocation status can be either \textit{Good} (indicating that the certificate is valid), \textit{Revoked} (indicating that the certificate has been
revoked for some reason), or \textit{Unknown} (indicating that the certificate is not known by the OCSP server and that is unable to give any answers about its status). Note that receiving a
positive answer does not mean that the certificate is still valid as it could be expired. Finally,
if a protocol error has occurred, the OCSP server answers with a message error. In this case, the response is not signed.
Below, we provide the list of those fields in an OCSP response affected by PQ transition:
\begin{itemize}
\item \textit{Responder ID} is the key hash of the OCSP responder that is used to identify the authority that signed the response.
\item \textit{Certificate ID} containing various fields: the issuer name hash, issuer key hash, and serial number of the certificate, as well as the hash algorithm used to calculate these hashes.
\item \textit{Signature Algorithm} contains the identifier of the algorithm used by the server to sign the response.
\item \textit{Signature} contains the signature computed on the hash of the DER encoding of OCSP response's ResponseData.
\end{itemize}

\section{PQ transition for PKI} \label{pq_transition_pki}


In this section, we discuss first the requirements at a high level, and then, we detail specific PQ-proposed formats for PQ-enabled X.509 certificates, CRLs, and OCSP, by presenting the challenges and implementation issues.

\subsection{Requirements} \label{generalrequirements}
We detail a set of requirements for PQ PKI and possible design choices for PQ transition. The requirements are both general and related to specific fields of the x.509 certificates, CRL, and OCSP protocol.

For a PQ certificate, the \textit{Subject Public Key Info} section (which contains the public key) must be permissive, therefore the Public Key Algorithm field must support PQ algorithms identifiers and the Public Key field must not have size limits, especially because PQ algorithms require bigger keys.

Moreover, the Issuer Key, namely the private key used by a CA to sign a certificate, must be strong since it plays a central role in establishing trust and ensuring the security and integrity of the PKI. Additionally, the key size is relevant so that it is possible to have small certificate chains. On the other side, the signature algorithm's speed is not relevant for offline operation but must be considered if certificates are generated on the fly. The same rules should be applied to the CRL's Issuer Key, although the key size is less relevant than for the CA, assuming the CRLs are created offline.

We observe that, in general, the signature applied by the CA on the certificate must be strong and short to ensure that the certificate chain is not too long. Depending on the context, e.g., signatures attached to documents, the signature algorithm speed might not be relevant, but the size could be significant (because the certificate chain must not be too big). 

On the other hand, in online protocols, such as TLS or OCSP, 
%
the signature algorithm speed is crucial for the system's overall performance. Meanwhile, the signature size might not be relevant, since there is typically a short certificate chain.

In addition to technical requirements, it is necessary to provide security recommendations too \cite{towards_security_recommendations_pq_pki}, like cryptographic agility \cite{BSI} \cite{pki_in_pq_era}, namely the ability of a security system to be able to rapidly switch between encryption mechanisms without affecting significantly the system's infrastructure, or cryptography in place meaning that there must be universal security requirements for the SDKs used by all the parties involved in PKI.


\subsection{Transition to PQ-based X.509 certificates} \label{PQX509}

There are currently three main methods/proposals to integrate PQC in digital certificates:

1. \textit{Quantum-safe certificates}: 
In the case of X.509 certificates, the simplest transition to use PQ
algorithms would be to put the PQ public keys directly into the existing fields of the \texttt{tbsCertificate} element and append the PQ signatures. For this type,
standardization of a new OID is needed without changes in the standard. However, PQ certificates can only be used with PQ applications, thus being a feasible solution only once all the involved systems are upgraded.


2. \textit{Hybrid certificates}: 
These certificates integrate PQ keys, algorithms, and signatures in \texttt{Subject Alt Public Key Info}, \texttt{Alt Signature Algorithm}, and \texttt{Alt Signature Value} (introduced in \cite{isoiec}) fields as non-critical extensions of the X.509v3 format. Meanwhile, classical algorithms are still kept in the \texttt{tbsCertificate} element, thus providing two or more signatures and keys in the same certificate.
This is generally considered a good option until the PQ is widely adopted.
As the format of a dual signature is out of scope of the NIST drafts, it is up to the application to specify how to parse signatures and verify them separately.


3. \textit{Composite certificates}: 
An alternative format proposed is not to use additional X.509v3 extensions, but to concatenate multiple cryptography algorithms in sequence to form a single key, signature algorithm, or signature value such that they can be used as a drop-in replacement for existing X.509 fields. The goal of composite certificates is to address the concern that neither the traditional algorithms
in use nor the PQC algorithms to be used are fully trusted. Implementing multi-key cryptographic operation based on composite certificates provides strong protection that breaking it requires breaking each of the component algorithms individually.


4. \textit{Parallel certificate chains}: 
An end-entity has two or more certificates with the same identity 
, but different public keys and signature algorithms (both classical and PQ). In this case, the burden is left on the application (relying party) that must select the proper certificate (chain) depending on the context at the strength required.

\subsection{Challenges for transition to PQ X.509 certificates}

We discuss here the implications (in terms of size, lifetime, and implementation aspects) that must be considered in the transition to PQ X.509 certificates.

\subsubsection{Size} 

Kampanakis et al. \cite{viability_pqx509} provides an insight on the practical consequences of a PQ implementation of X.509 certificates in terms of size.
Since the size of the public key and signatures can grow to many \SI{}{\kilo\byte}, this could affect applications and protocols that use them in X.509 certificates.
The effect is a higher transmission overhead which leads to delays in connection setup, IP fragmentation, and wasted bandwidth for connections that transfer small amounts of data.

To cope with these problems, application protocols can be adapted for example several mechanisms are already in use in TLS to handle certificate size issues, such as fragmentation for chains longer than \SI{16}{\kilo\byte}, client caching (the server can avoid resending a certificate or chain if it
is already cached by the client), or compression.


\subsubsection{Certificate lifetime and criticality}
Quantum computing will also affect certificates based on their expiration: it is important to prioritize the transition for long-lived (and highly critical) certificates like root CA certificates (with validity longer than 10 years) rather than the medium or short-lived ones (with validity of 3 months up to one year), which are used to authenticate for example web servers within TLS.

\subsubsection{Implementation aspects}

Several works have addressed the implementation aspects of transitioning to PQ PKI \cite{viability_pqx509} \cite{transitioning_to_quantum_resistant_pki}
\cite{integration_of_qs_algo_in_x509v3}
\cite{performance_pq_digital_certificates}
\cite{impact_pq_hybrid_on_pki}, and provide useful results.
For instance, experiments with TLS libraries and web browsers have been conducted with hybrid certificates, where the PQ material (key, algorithm) is put inside non-critical extensions. Some size limitations have been observed for specific TLS libraries. For example, in mbedTLS 2.4.2 the certificate cannot exceed \SI{9}{\kilo\byte}, while OpenSSL 1.0.2 can handle up to \SI{43}{\kilo\byte} certificates (size of a SPHINCS+ certificate).
The lesson learned is that backward compatibility is most easily maintained when PQ objects can be placed as non-critical certificate extensions.



Moreover, currently there are already some available implementations and simulations of PQ PKI, that can be divided in proprietary and open source solutions.
Certified Security Solutions (CSS) and ISARA introduced the ``First and Only Quantum-Safe, Full Stack PKI''\footnote{\url{https://www.keyfactor.com/press-releases/css-and-isara-introduce-the-first-and-only-quantum-safe-full-stack-pki/}} as a solution for the automotive industry. 
In addition, different open-source projects contribute to the PQ PKI landscape, such as the \textit{libpqcrypto}\footnote{\url{https://libpqcrypto.org/index.html}} PQC library, which offers an implementation of PQ algorithms 
designed to be resistant to quantum attacks. It includes implementations of several lattice-based, code-based, and other quantum-resistant algorithms.
Another implementation is given in 
\cite{implementation_x509-compliant_qs}, which is a proof-of-concept implementation of a PQ PKI, using a forked version of OpenSSL 
     integrated with the Open Quantum Safe (OQS) library\footnote{\url{https://github.com/open-quantum-safe/liboqs}}. This implementation employs a composite approach, combining ECDSA and Dilithium into the same X.509 fields by concatenating the shared secrets.
Finally, we mention GlobalSign's repository holding x509 objects with PQ algorithms\footnote{\url{https://github.com/globalsign/example-pq-safe-x509}}. The repository contains implementations of certificates, CRLs, and OCSP. Moreover, \cite{pq_certificates_examples} addresses best practices for implementing PQ X.509 certificates.


\subsection{Design} \label{design}
Based on the requirements in Sect.~\ref{generalrequirements}, we propose a design choice for transitioning to PQ PKI. 
The first step involves implementing hybrid certificates containing both classical and PQ algorithms within the same certificate. This approach offers a bridge between existing and quantum-resistant cryptography.

The next stage in the roadmap is to implement parallel certificate chains. This structure provides flexibility with minimal impact on the certificate chain's size. Parallel certificates feature multiple certificates for the same entity, using different cryptographic algorithms. This approach allows for interoperability and backward compatibility with classical systems, while introducing PQ security.

Composite certificates provide the best solution for higher security levels, especially Root CAs. This structure requires an attacker to compromise both classical and PQ components to break the certificate. However, composite certificates also increase the chain's complexity, affecting the verification process, and generally result in larger certificate sizes.

Lastly, the final goal is to transition to pure PQ certificates, representing a complete migration to quantum-resistant cryptography. This stage should occur once all relevant systems have been upgraded to support PQ algorithms.

 About the certificate format, it is already indicated in \cite{BSI} that certificate extensions will be used to accommodate the new signature schemes.
 Specifically, extensions like \texttt{subjectAltPublicKeyInfo}, \texttt{altSignatureAlgorithm}, and \texttt{altSignatureValue} are defined to support alternative public keys and cryptographic algorithms. These are intended as a transitional solution, allowing a smooth migration to quantum-safe signature schemes.
 
For Issuer Keys, which are critical to the security of PKI as explained in Sect. \ref{generalrequirements}, we recommend using algorithms with a security level of 5, such as Falcon 1024, Dilithium 5, or SPHINCS+ SLH-DSA-256. Among these, Falcon 1024 is a strong choice due to its relatively short signature size, despite not having the shortest private key. Research by Raavi et al. \cite{performance_pq_digital_certificates} suggests that Falcon algorithms are ideal when certificate size is the primary concern, while Dilithium algorithms are better suited for scenarios where fast verification is critical, such as OCSP protocol. Meanwhile, \cite{BSI} considers hash-based signature schemes like LMS and XMSS suitable for long-lived root certificates but less ideal for end-user certificates due to their stateful nature. Therefore, hash-based algorithms could be useful for building a mixed PKI, where different signature schemes are used for root certificates compared to end-user certificates.

\subsection{Transition to PQ-based CRL and OCSP}

The CRLs will also have to be updated (in each transition stage) to include PQ (hybrid, composite, or pure) certificates. Otherwise, malicious parties could try to falsely prove that valid certificates have been revoked or erase the revocation of certificates that should no longer be trusted.

Again, the signature of the PQ-based CRL will be bigger than the current CRLs with RSA/ECC keys. However, the increased size of the CRL is unlikely to cause significant issues, because CRLs can become quite large when the issuing CA has processed many revocations.
It is rather a problem of adapting the format of the CRL with additional extensions. We can foresee pure PQ CRLs signed only with PQ keys and algorithm, hybrid CRLs that contain both a traditional (e.g. RSA) signature and a PQ signature, or a composite CRL signed by a CA with a composite X.509 certificate.


Experiments performed with hybrid certificates in OCSP are given in \cite{impact_pq_hybrid_on_pki} both for OCSP (client and server) executed with command line tools and with OCSP integrated into browsers. 
In the abovementioned research paper, the client creates an OCSP request, which could exploit a hybrid certificate to sign the request (step 1). Then the client sends the OCSP request to the OCSP responder (step 2), which will create an OCSP response signed with its hybrid certificate (step 3). Finally, the OCSP server sends back the OCSP response to the client (step 4). All the signatures used only the entity's RSA key but the messages sent in steps 2 and 4 contained certificates with PQ extensions, implying that they were larger than the messages in the classical OCSP protocol.

The authors studied the impact of certificate size on both the OCSP client and the OCSP responder, tested with OpenSSL 1.0.2-fips 26 January 2017/1.1.1b 26 February 2019 and CFSSL 1.3.2 as command line OCSP servers on CentOS 7, while OpenSSL 1.0.2p of 14 August 2018 was used as an OCSP command line on Windows 8.1 Enterprise. They have tested hybrid certificates of various types and sizes, such as 'S' (\SI{49216}{\byte} bytes PQ signature, and a \SI{50434}{\byte} certificate), 'P' (\SI{209478}{\byte} PQ signature, and a certificate size of \SI{210692}{\byte}). and 'G' (\SI{104}{\byte} PQ signature, and \SI{3605052}{\byte} certificate).
In terms of functionality, they obtained good results because the OCSP responders worked for all sizes of hybrid certificates. Moreover, the OCSP server worked for the OCSP requests with or without a signature signed by the client using its 'conventional' RSA key stored in a hybrid certificate.
%
Unfortunately, the paper does not indicate the time spent to perform the OCSP transaction. The OCSP processing time is relevant in some application contexts, and we have identified it as a main general requirement in Sect. \ref{generalrequirements}.
The same authors have also tested web OCSP implementations, in testbed setups using  Mozilla Firefox and Internet Explorer web browsers.
The authors did not test on Google Chrome because OCSP checks have been disabled in recent versions, for non-Extended Validation certificates, as indicated in \cite{BerbecaruIEEEAccess}.
OCSP checking worked correctly in IE for all sizes of OCSP responses, with some exceptions.
Unfortunately, also in this case no indication is given on the delay perceived by the user due to the processing of bigger size (hybrid) certificates.

\section{Conclusions and Future Work} \label{conclusions}

It is known that the PQ transition must be done soon, but the question is how long it will take this process and how complex would be the integration of PQ keys and signatures into security protocols and formats used today.
We presented briefly the algorithms and the current status of NIST's standardization process. Then, we focused on the main challenges and requirements for integrating PQ into X.509 certificates, identifying the main fields and extensions affected by the PQ transition. This investigation should be improved by including more specific case studies demonstrating the implementation in real-world systems. We delineate a roadmap for PQ transition, specifying which certificate formats are more suitable in each stage. We discuss challenges in terms of size, speed, and implementation choices. Finally, we touch PQ transition for standard revocation mechanisms, i.e., CRL and OCSP.
Future work could address PKI implementation measurements with indicated libraries and common browsers in laboratory testbed, enhancements to ensure efficient certificate verification, 
and the implications of PQ transition for the Certificate Transparency ecosystem.
%

{\footnotesize
{\textbf{Acknowledgments.}
This work has been developed within the QUBIP European Project ({\url{https://qubip.eu/}}), funded by the European Union under the Horizon Europe framework programme [grant agreement no. 101119746]. Dr. Diana Gratiela Berbecaru carried out her work within the Ministerial Decree no. 1062/2021 and received funding from the FSE REACT-EU - PON Ricerca e Innovazione 2014-2020.}


\end{document}